\documentclass[aps,twocolumn,groupedaddress,showpacs]{revtex4}
\usepackage{graphicx}%

\begin{document}
\bibliographystyle{apsrev}


\title[nanotube epi-fringes]{Lattice fringe signatures of epitaxy on nanotubes}


\author{Jinfeng Wang}
\author{P. Fraundorf}
\email[]{pfraundorf@umsl.edu}
\affiliation{Physics \& Astronomy and Center for Molecular Electronics, U. of Missouri-StL (63121), St. Louis, MO, USA}

\author{Yangchuan Xing}
\affiliation{Chemical \& Biological Engineering, U. Missouri-R (65409), Rolla, MO, USA}


\date{\today}

\begin{abstract}
Carbon nanotubes are of potential interest as heterogeneous catalysis supports, in part because they offer a high surface area hexagonal array of carbon atoms for columnar or epitaxial attachment.  Fringe visibility modeling of electron microscope lattice images allows one to investigate the relationship between individual nanoparticles and such nanotube supports. We show specifically how (111) columnar or epitaxial growth of FCC metal lattices, on carbon nanotubes viewed side-on, results in well-defined patterns of (111)-fringe orientations with respect to the tube axis.  In the epitaxial case, the 
observations also provide information on chirality of the nanotube's outermost graphene sheet.
\end{abstract}
\pacs{03.30.+p, 01.40.Gm, 01.55.+b}

\maketitle

\tableofcontents
\section{Introduction}

Carbon nanotubes are of increasing interest for nanotechnology 
applications in general \cite{Dresselhaus96, Dai02}.  
They are also of interest as supports 
for heterogeneous catalysts \cite{Rylander79, Planeix94}, 
for example in fuel cell 
applications \cite{Li02}.  Carbon nanotubes also have advantages for 
electron microscopy, because their near-cylindrical symmetry 
and propensity of lie perpendicular to the electron beam 
makes data acquisition and interpretation much simpler than 
for single crystal supports.  

\begin{figure}
\includegraphics[scale=.83]{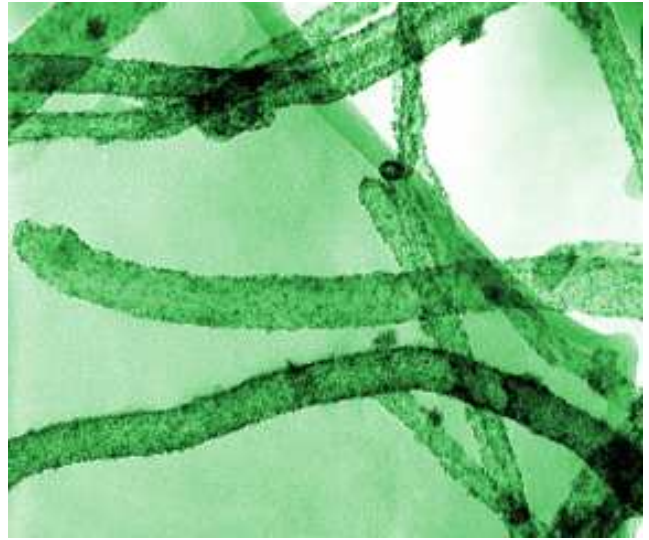}
\caption{Low magnification TEM image of Pt 
particles supported by carbon nanotubes \cite{Xing04}.}
\label{Fig1}
\end{figure}

We show here that, with help from recent 
work on the theory of lattice fringe visibility \cite{wqpfx.lptt, Fraundorf05JAP}, 
that the crystallographic relationship between catalyst nanocrystals 
and their nanotube supports (Fig. \ref{Fig1}) can sometimes be inferred from a 
single lattice fringe image.  If the relationship is epitaxial, 
one might conversely decorate tubes in order to determine the 
chirality of their outer sheet.  We illustrate the approach 
with a specific case:  Growth of FCC metals on the top (0002) graphite 
(or graphene) surface of a carbon nanotube.  The theory is 
easily extended to growth of other 
lattice types.  Application to experimental images 
of Pt fuel cell catalyst crystals, on carbon nanotube supports, 
illustrates the strategy.

\section{Theory}

\begin{figure}
\includegraphics[scale=.8]{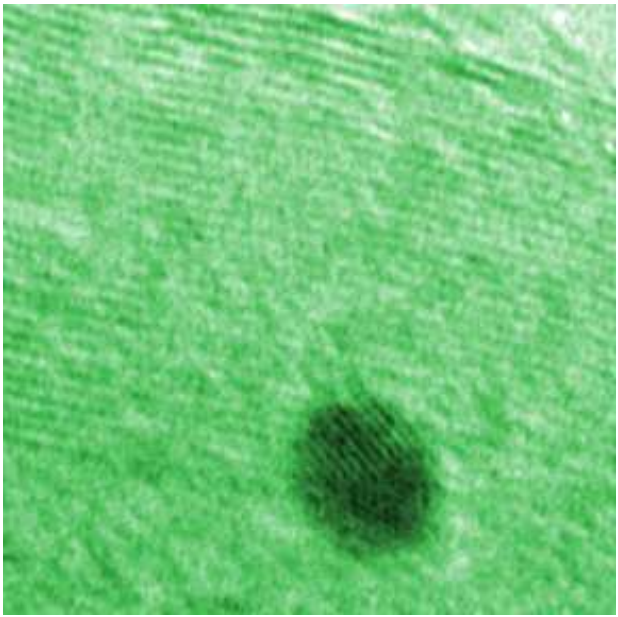}
\caption{High resolution image of a bicrystalline 
Pt nanoparticle on a carbon nanotube, half of which 
shows clear 0.26 nm (111) fringes.}
\label{Fig2}
\end{figure}

As aberration-correction makes it possible for electron 
microscope images to provide resolutions into the subAngstrom 
range, quantitative information contained in lattice 
fringes (Fig. \ref{Fig2}) will become increasingly robust \cite{Fraundorf05JAP}.  
In particular, on tilting away from the edge-on 
view of a lattice-plane in the transmission electron 
microscope (TEM), one encounters a range of 
incident electron angles (e.g. relative to lattice-plane 
parallel) within which the lattice plane's reciprocal 
lattice spots continue to intersect the Ewald sphere.  
Hence lattice fringes associated with those planes remain visible. 
The upper bound of this "visiblity band half-angle" (with 
the largest term first in the ``thin specimen'' limit) is
\begin{equation}
\alpha = \sin^{-1} \left[ d \frac{\Gamma}{t} + \frac{\lambda}{2 d} \left( 1 - 
\left(d \frac{\Gamma}{t} \right)^2 \right) \right] .
\label{AlphaMax}
\end{equation}
Here $d$ is the spacing of lattice planes, $t$ is the crystal thickness, $\lambda$ is the wavelength of the electrons, and $\Gamma$ is a ``visibility factor'' on the order of 1 that empirically accounts for the signal-to-noise in the method used to detect fringes.  The effective radius of the reciprocal lattice spot in this model (or excitation 
error $s$ at which the fringe fades to background) is $\Gamma/t$.

\begin{figure}
\includegraphics[scale=.6]{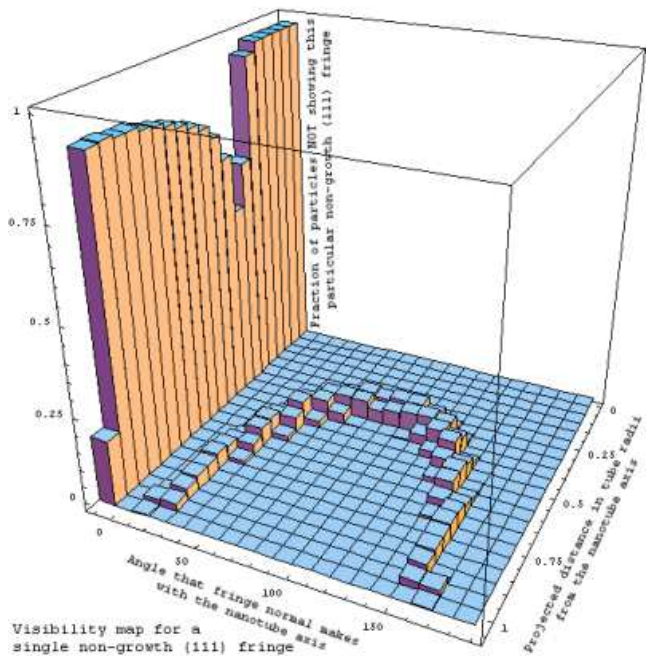}
\caption{Fraction of (111) columnar nanoparticles showing a given non-growth 
(111) fringe, as a function of azimuth angle of the fringe, and projected fractional radius, from the nanotube axis.  This plot is for particles of 4.8 nm thickness.  The histogram on the left roughly estimates the fraction of particles NOT showing this fringe.}
\label{Fig3}
\end{figure}

We apply this theory to three types of relationship between 
supported FCC nanocrystals and underlying carbon and/or BN nanotubes: 
randomly oriented, (111) columnar, and (111) epitaxy.  The objective 
is to illustrate the distribution of (111) lattice fringe orientations expected 
on a nanotube encountered by the electron beam ``side on''.  The 
result is expected to be insensitive to slight deviations of 
the beam from perpendicularity to the tube axis. 

\subsection{Random orientations}

  For the random orientation case, we  
let each nanoparticle orient randomly to the 
electron beam.  This might occur, for example, 
if the nanoparticles grow or accumulate with 
no regard to the underlying surface.  Each 
nanoparticle has four sets of (111) 
lattice planes, in tetrahedral orientation to 
each other.  From equation \ref{AlphaMax}, 
the fringe will show up in the center of a 
28[nm] diameter particle only when the electron 
beam is tilted by less than an $\alpha$ value 
of about 5 degrees out of 
that plane.  For sufficiently thin particles, 
this result is relatively insensitive to the 
high energy of the electrons (e.g. 300[kV]) 
that the microscope is using.

Given that FCC crystals have four (111) 
planes which pairwise intersect down each of six 
$\left\langle 110 \right\rangle$ beam directions at an angle of 
$\theta_g = \cos^{-1}(1/3) \cong 70.5$ degrees, the 
fraction $f_2$ of particles showing cross-fringes 
at their center will be \cite{Fraundorf05JAP} 
about six times $2 \alpha^2 / (\pi \sin(\theta_g))$ 
or around 3\%.  The fraction $f_1$ of 
particles showing only 1 fringe at center will be 
$4 \sin(\alpha) - 2 f_2$ or around 29\%.  
Hence about 68\% of the particles 
will show no central fringes.
Most importantly, as shown in
the top central panel of \ref{Fig5}, these 
probabilities will be independent of the 
projected position of each nanoparticle with 
respect to the nanotube's axis.

\subsection{Columnar growth down (111)}

  If the nanoparticles systematically 
grow upward from their substrate, 
along a column perpendicular to  
a (111) ``growth plane'' with no regard 
to the azimuthal orientation of the 
substrate beneath, we refer to this 
as (111) columnar growth.  In this case
the fringe abundances remain the same, 
but now those abundances depend on the 
projected position of the nanoparticle 
with respect to the nanotube axis.

  In particular, the surface normal 
of a nanotube is tilted with respect 
to the incident electron beam by 
an angle $\theta$ equal to $\sin^{-1}(r/R)$, 
where r is the distance of the particle 
from the projected nanotube tube axis, and R is
the nanotube radius.  Consider now the 
unit normal to a lattice plane tilted up from 
the surface normal of the nanotube by $\theta_g$, 
on an axis which makes an angle of $\phi$ 
with respect to the tube axis.  
If we let the electron beam direction define 
our viewer z-axis, and the projected tube-axis define 
the viewer x-axis, then that lattice plane normal in
viewer coordinates has
x-component $\cos(\phi) \sin(\theta_g)$, 
y-component 
$\cos(\theta_g) \sin(\theta) + \sin(\theta_g) \sin(\phi) \cos(\theta)$, 
and z-component 
$\cos(\theta_g) \cos(\theta) - \sin(\theta_g) \sin(\phi) \sin(\theta)$.
As long as the normal to this plane deviates 
from perpendicularity to the beam by less than 
$\alpha$, the fringe is visible and makes the 
angle predicted by it's x and y components.

\begin{figure}
\includegraphics[scale=.8]{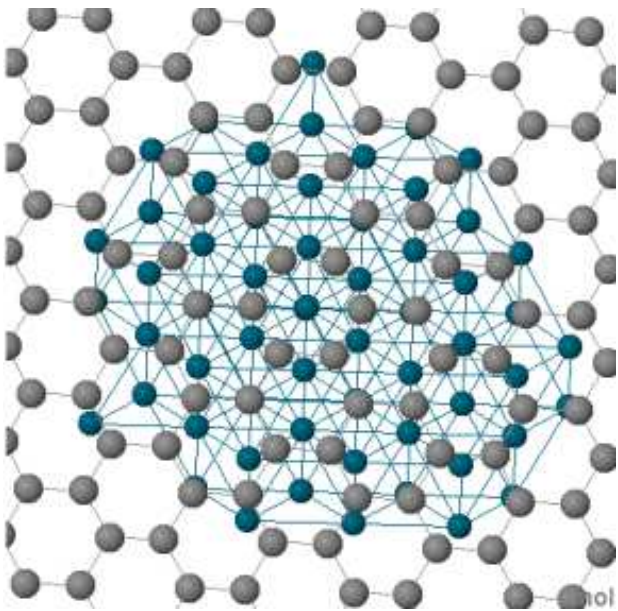}
\caption{Model of Pd (65 atoms) epitaxially grown 
on graphite (0002) or a graphene sheet surface.}
\label{Fig4}
\end{figure}

  Thus the growth (111) planes 
are only visible on top and bottom 
sides of the nanotube, when the beam 
direction is within three degrees of 
tangent to the nanotube surface.  Also, 
as shown in Fig. \ref{Fig3}, no 
particle-center (111) fringes are 
visible at all for projected 
locations on the nanotube within 
approximately one 
third of the tube radius from the 
center.  The
upper right corner of Fig. \ref{Fig5} 
shows that the resulting distribution 
of fringes is easily distinguishable 
from that expected for 
randomly-oriented particles.

\subsection{Epitaxial growth on (111)}

According to diffraction data in the literature \cite{Humbert91}, 
epitaxy of FCC Pd on single crystal graphite (0002) planes 
typically takes place 
with $(111)_{Pd}$ parallel to $(0002)_{graphite}$ in 
the growth direction, 
and $[112]_{Pd}$ parallel 
to $[01.0]_{graphene}$ in the plane of the substrate.  
This fully specifies the expected fringe 
distributions.  

\begin{figure}
\includegraphics[scale=.7]{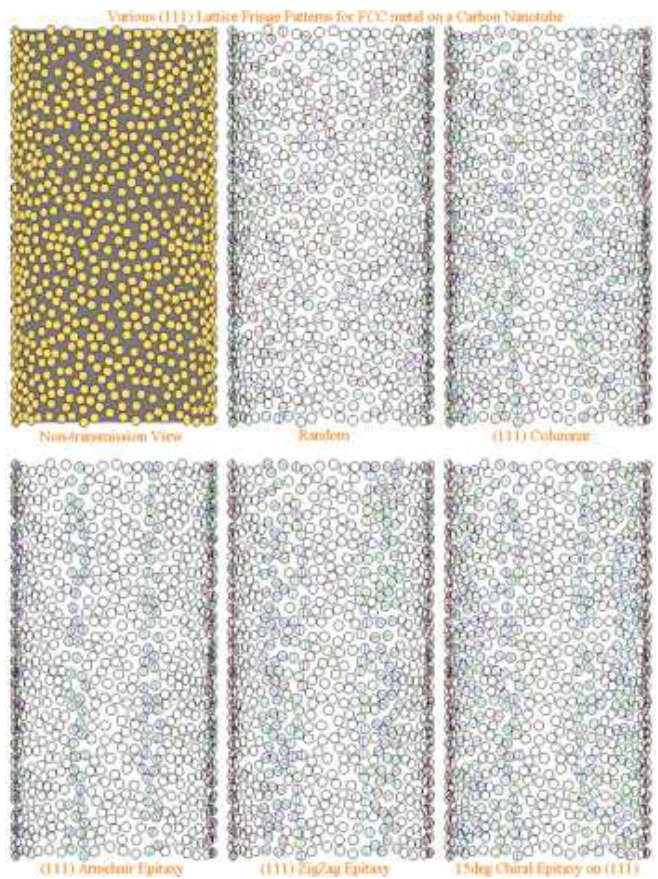}
\caption{Illustrated (111) fringe patterns expected 
for various 28A particles grown on a carbon 
nanotube, and observed with 300kV electrons.}
\label{Fig5}
\end{figure}

The bottom up view of such a 65 Pd-atom 
nanocrystal on graphene is shown in Fig. \ref{Fig4}.
Here for simplicity we've approximately positioned 
two of every three bottom-layer Pd atoms directly over a carbon 
atom, with the result that one of every three is 
over the center of a cyclohex graphene ring.  The 
actual transitional alignment of the nanoparticle 
should have little effect on metal fringe visibility, 
provided that contrast from the underlying carbon 
does not obscure the metal lattice.

Because the Pd lattice azimuth is now fixed 
with respect to top graphene sheet of the underlying 
nanotube, the fringe visible will be fully 
specified by the projected position for a given 
top sheet chirality.  The result is illustrated 
at the bottom of Fig. \ref{Fig5} for armchair, 
zigzag, and ``15 degree'' chiral top-layer 
orientations, in sequence.  As a fringe 
benefit, epitaxial ``decoration'' of carbon 
nanotubes might allow routine 
determination of top-layer tube chirality.  Of 
course, as we discuss below, preparation of 
nanotube surfaces sufficiently clean and flat
for epitaxial decoration may not be trivial.

\section{Experimental Test}

In this experiment, we use a Philips EM430 supertwin TEM with point resolution near 0.2 nm to image Pt nanoparticles on nanotubes. In the images, we can find many Pt particle fringes. Pt is one of FCC elements [1].  Columnar growth along FCC (111) is, among other things, expected to result in (111) fringes parallel to the tube axis only along the tube edge, and at a projected distance from each tube edge of 1/3 the tube radius.   When we magnify the fringe images, we find instead that projected Pt fringe angles are randomly distributed on the Carbon nanotube. So in this case nanoparticle growth is not along (111) columns.


\section{Discussion}

The foregoing illustrates the particular ease 
with which preferential crystallographic orientation 
can be detected in particles on nanotubes.  
However, preferential orientation on carbon 
nanotubes is complicated by two factors.  

The first is surface integrity.  For example 
in the experimental specimens above, it is 
easy to see that intact graphene layers only 
make it within a few atomic layers of the 
surface.  This is likely due to an intentional 
roughening treatment designed to improve 
particle adhesion. Hence epitaxial growth would be 
impossible on these specimens, although 
columnar growth still conceivable.  

The second complication is surface curvature.  
We're not aware of experimental data on the 
effects of graphene sheet curvature on epitaxy.  
For example, it may be possible only on 
larger diameter tubes.  On the other hand, 
the technique of detecting epitaxy suggests 
that nanotubes substrates serve as an excellent candidates 
for experimental investigation of these sheet 
curvature effects.  

Finally, we've worked through 
the mathematics here of one simple, and 
common, example of possible coherent relationships 
between carbon nanotubes and adhering particles.  
The methods used here should work seamlessly 
in other cases as well.



\begin{acknowledgments}
Thanks to MEMC Electronic Materials Company and 
Monsanto Corporation for regional facility support.
\end{acknowledgments}

\bibliography{temr1wq2}


\end{document}